# IMPACT OF A TREATMENT AS PREVENTION STRATEGY ON HEPATITIS C VIRUS TRANSMISSION AND ON MORBIDITY IN PEOPLE WHO INJECT DRUGS

Anthony Cousien[,1,2], Viet Chi Tran[3], Sylvie Deuffic-Burban[1,2,4], Marie Jauffret-Roustide[5,6], Jean-Stéphane Dhersin[7], Yazdan Yazdanpanah[1,2,8]

[1]IAME, UMR 1137, INSERM, F-75018 Paris, France

[2]IAME, UMR 1137, Univ Paris Diderot, Sorbonne Paris Cité, F-75018 Paris, France

[3]Laboratoire Paul Painlevé UMR CNRS 8524, UFR de Mathématiques, Université des Sciences et Technologies Lille 1, Cité Scientifique, Villeneuve d'Ascq, France

[4]Inserm, LIRIC-UMR995, F-59000 Lille, France; Univ Lille, F-59000 Lille, France

[5]CERMES3: Centre de Recherche Médecine, Sciences, Santé, Santé Mentale et Société, (INSERM U988/UMR CNRS8211/Université Paris Descartes, Ecole des Hautes Etudes en Sciences Sociales), Paris, France

[6] Institut de Veille Sanitaire, Saint-Maurice, France

[7]Université Paris 13, Sorbonne Paris Cité, LAGA, CNRS, UMR 7539, F-93430, Villetaneuse, France

[8]Service des Maladies Infectieuses et Tropicales, Hôpital Bichat Claude Bernard, Paris, France

**Contact information**

Université Paris 7, UFR de Médecine - Site Bichat

Inserm UMR1137, IAME

16 rue Henri Huchard - 75018 Paris –France

Email: anthony.cousien@inserm.fr

**List of abbreviations**

HCV: hepatitis C virus

PWID: people who inject drugs

DAA: direct-acting antiviral

SVR: sustained virological response

HCC: hepatocellular carcinoma

LTC: linkage to care

**Financial support**

This study was funded by the French AgenceNationale de Recherchesur le Sidaet les Hépatitesvirales (ANRS, http://www.anrs.fr), grant number 95146. Viet Chi Tran was supported in part by the Labex CEMPI (ANR-11-LABX-0007-01).

**ABSTRACT**

Background: Highly effective direct-acting antiviral (DAA) regimens (90% efficacy) are becoming available for hepatitis C virus (HCV) treatment. This therapeutic revolution leads us to consider possibility of eradicating the virus. However, for this, an effective cascade of care is required.

Methods: In the context of the incoming DAAs, we used a dynamic individual-based model including a model of the people who inject drugs (PWID) social network to simulate the impact of improved testing, linkage to care, and adherence to treatment, and of modified treatment recommendation on the transmission and on the morbidity of HCV in PWID in France.

Results: Under the current incidence and cascade of care, with treatment initiated at fibrosis stage ≥F2, the HCV prevalence decreased from 42.8% to 24.9% [95% confidence interval 24.8%–24.9%] after 10 years. Changing treatment initiation criteria to treat from F0 was the only intervention leading to a substantial additional decrease in the prevalence, which fell to 11.6% [11.6%–11.7%] at 10 years. Combining this change with improved testing, linkage to care, and adherence to treatment decreased HCV prevalence to 7% [7%–7.1%] at 10 years and avoided 15.3% [14.0%-16.6%] and 29.0% [27.9%–30.1%] of cirrhosis complications over 10 and 40 years respectively.

Conclusion: A high decrease in viral transmission occurs only when treatment is initiated before liver disease progresses to severe stages, suggesting that systematic treatment in PWID, where incidence remains high, would be beneficial. However, eradication will be difficult to achieve.

# INTRODUCTION

Hepatitis C Virus (HCV) is responsible for more than 350,000 deaths every year worldwide (1). In high income countries, the main HCV transmission route is injection drug use(2). In these countries, despite the introduction since the 80s of harm reduction measures, HCV prevalence among people who inject drugs (PWID) remains high,often above 60% (3).

Until recently,HCV treatment standard of care was a dual therapy combiningpeg-interferon and ribavirin that was moderately effective and with a high proportion of adverse events (4, 5). New direct-acting antiviral (DAA) based interferon-freeregimens, associated with highersustained virological response (SVR) rates, better tolerance profiles, and shorter durations arenowavailable(6-11). These therapeutic improvements raise the question of usingHCV treatment as a mean of preventing HCV transmission in PWID, and, in the longer-term of possibly eradicating the virus. However, the effectiveness of this strategy depends on several factors: time to diagnosis, patients' linkage to care, and treatment initiation criteriato achieve early treatment initiation; and adherence to care and treatment, to achieve a high SVR rate.

Evaluating the impact of incoming treatments or improvementsin the cascade of care on transmission inPWIDthrough traditional epidemiological studies faces issues related to feasibility and costs. Dynamic modeling of HCV transmission inPWID is an interesting alternative allowing us to estimate the impact of various scenarioson the spread of HCV in this population. Numerous models have been proposed in the literature(12). However, most of them did not considered the social network of PWID, which impactsviral transmission (13); and they have often estimated strategies impacting only a specific point in the cascade of care (and in particular treatment rate or treatment effectiveness), not the overall cascade.

The aim of this analysis wasto estimate, in the context of incoming DAAs regimens, the impact of HCV testing, linkage to care, and treatment efficiency improvement, andof changes in treatment initiation criteria,on HCV transmission andHCV-related morbidityinPWID. For this purpose, we useda dynamic stochastic individual-based model (IBM) with a social network model of PWID and a natural history model of chronic hepatitis C.

## METHODS

Here, we present the simulated population, the structure of the model, details of the different scenarios we evaluated, and the sensitivity analyses we performed. More details regarding the hypotheses we made on parameter values are given in Supplementary Information S1.

**Population**

In this analysis, we focused on a population with the characteristics of the PWID population in France. We considered that the size of the population was constant at 10,000 PWID – the order of magnitude of the drug user population in the main cities of France(14). Each PWID is characterized by 1) gender, to take into account differences of mortality between men and women(15): the gender of each new PWID is drawn following a probability $p_M$ of being a man; 2) a set of injecting partners: we draw for each PWID a group of other PWID in the population according to a random graph model (see Supplementary Information S1) susceptible of infecting or becoming infected by the index PWID; 3) his/her status relating to injection: current or former injector, i.e. injector after cessation of drug use; 4) his/her HCV-infection status; 5) if HCV-infected, his/her status regarding HCV infection knowledge and linkage to care, liver fibrosis stage, and treatment.

**Model**

*Social network of injecting partners model*

One of our objectives was to simulate possible pathways of HCV transmission in PWID. HCV is mainly transmitted by needles/syringes sharing in the PWID population; however paraphernalia sharing (e.g. filter, spoon) seems to play an important role too (16). To take into account the global risk of infection for a PWID, we chose, as previously described by Rolls *et al.*, to model the network of the sharing partners: two PWID are linked together if they inject together even without sharing needles/syringes (13).

The mathematical literature describes numerous models of networks (17). In the case of the injecting partners' network, the choice of a graph model is difficult, because no data are currently available on PWID social networks in France. Therefore, we used an Erdős-Rényi model where each PWID

"couple" is linked with a fixed probability $p$ (18). This model is simple to calibrate with only one parameter, which can easily be obtained with field studies: knowing $\bar{d}$, the mean number of injecting partners per PWID (i.e. the *degree*), and $N$, the size of the population, the parameter of the Erdös-Rényi model is $p = \bar{d}/N$.

We assumed a static network, i.e. where there is no change in the links between PWID overtime. In addition, we hypothesized that each PWID who dies would be replaced by a new PWID.

*Transmission and care model*

We divided the population into two states: current injection injector and former injector. Each new PWID in the population is a current injector, and we attributed each one a duration for their injecting career, drawn from an exponential distribution of mean $1/\theta$, where $\theta$ is the rate of drug use cessation. Only current injectors can transmit HCV or be infected with HCV; we assume that sexual transmission of HCV in this population is negligible, as the incidence rate is approximately one per 190,000 sexual contacts for heterosexual relationships (19). This mode of transmission was therefore not considered. Figure 1 describes the possible states through which PWID can progress. Briefly, new PWID start in S (Susceptible at high risk of infection) for an average duration of one year, as there is a large amount of data showing that PWID are at higher risk of HCV infection during the first year of their injecting career (20). After a time determined by the rate $\eta$, they progress to S' (Susceptible at low risk of infection), in which their risk of infection is lower. In S (respectively S'), each PWID $k$ may be infected at rate $\beta I(k)$ (respectively $\beta' I(k)$), with $\beta$ (respectively $\beta'$) the infection rate per infected partner, and $I(k)$ the PWID's number of infectious injecting partners. After infection, PWID progress to stage A (Acute hepatitis C), in which they stay for a fixed time $T_a$. Acute Hepatitis C can lead to a spontaneous recovery with a probability $p_r$, and the PWID returns in S'. Given the length of time between the beginning of the injecting career and the end of the acute hepatitis C, we did not consider transitions between A and S. If spontaneous recovery does not occur, PWID progress from A to UC (Undiagnosed Chronic hepatitis C) with a probability equal to $1 - p_r$. A chronic hepatitis C is diagnosed at rate $\delta$, which depends on the status of current/former injectors. When a PWID is diagnosed, he/she progress to DNLC (Diagnosed and Non-Linked to care Chronic hepatitis C). A PWID is considered linked to care if he/she has one or more contact per year with a health care service regarding his/her HCV-infection. When a PWID is linked to care with a rate $\phi_{Link}$, he/she progress from DNLC to DLC (Diagnosed and Linked to care Chronic hepatitis C). However, he/she can be lost-to-follow at rate $\phi_{Lost}$, and in this case the PWID returns to the DLC status. We considered that PWID with cirrhosis complications (i.e. decompensated cirrhosis and/or hepatocellular carcinoma (HCC), see "Natural history" below) are always linked to care because of the severity of their illness.

PWID linked to care and who have a Metavir score between F2 and F4 (see subsection "Natural history model") are immediately treated for chronic hepatitis C, according to French national guidelines in December 2014 (21), and progress to T (Treatment). Treatment has a fixed duration $T_t$, and can lead to SVR with a probability $p_{SVR}$: in this case, the PWID returns to the S' state; he/she can be re-infected in the same manner as PWID who have never been infected with HCV. If the PWID does not respond to treatment he/she progress to the Non-SVR state with probability $1 - p_{SVR}$. This is an absorbing state, i.e. a PWID cannot escape this state, as we did not include the possibility of retreatment. The effectiveness of the DAAs in real-life and especially in PWID is not currently available. We therefore broke down $p_{SVR}$ into two variables such as $p_{SVR} = r \times e$, where $e$ is the treatment efficacy observed in clinical trials and $r$ is the ratio of the effectiveness of treatment in real-life to the efficacy in clinical trials.

In each state of our model, we applied a mortality rate for deaths unrelated to HCV infection (i.e. unrelated to a cirrhosis complication). This rate, denoted $\mu$, depends on two factors: the gender of the injector and the injector's status (current or former injector) (15). Mortality due to hepatitis C was taken into account in the natural history model described below.

*Natural history model*

The natural history model (Figure 2) describes the liver disease progression in HCV-infected PWID. It describes the fibrosis progression using the Metavir Score (22): F0 = no fibrosis, F1 = portal fibrosis without septa, F2 = portal fibrosis with few septa, F3 = numerous septa without cirrhosis, F4 = cirrhosis. In this analysis we grouped F0 and F1, and F2 and F3 in unique states (F0/F1; F2/F3).

The second part of the model describes cirrhosis complications. The first complication is the decompensated cirrhosis, and the second is HCC. In the model, decompensated cirrhosis may progress to HCC. Finally, those complications lead to death related to HCV-infection.

In case of successful treatment (see previous subsection), we considered that the PWID fibrosis regresses to F0, except if the patient has already developed cirrhosis (F4) (23). In this case, we considered that cirrhosis complications may still occur (23) with the same rate as that with chronic hepatitis C.

**Input Parameters**

Input parameters were mainly derived from ANRS-Coquelicotstudydata, which was a HCV-seroprevalencecross-sectional survey conducted among drug users in France (24), and from the medical literature. Missing parameters were fitted by Approximate Bayesian Computation (ABC – see Supplementary Information S2)(25).

A summary of parameters and their default values are given in Supplementary Information S1, Table S1.

**Tested scenarios**

We simulated 7 scenarios, corresponding to different testing rates, linkage to care and loss to follow-up rates, treatment initiation criteria or SVR rates(see Table 1).

*Scenario 1* (reference scenario) is the scenario with the parameter values presented in Table S1 in Supplementary Information, corresponding to the current cascade of care in the French PWID population. In the baseline analysis, for the nature of HCV treatment,we considered that patients receivedDAAs-based regimens. SVR for these regimens are estimated at$e$=90% in clinical trials and treatment duration at $T_t$=12 weeks (6-11). However, we decreased the efficacy rate of these treatments since they were derived from clinical trials in mostly non-IDUs population. To make these modifications, we applied the coefficient $r$ derived from the ratio of the SVR inreal life to the SVR in clinical trials for peg-interferon + ribavirin($r$=0.903)(26).

*Scenario 2* is scenario 1 with an improved testing rate. The mean time between infection and testing ($1/\theta$) is considered to be 0.5 years for all individuals (vs. 1.25/1.45 years for current/former PWID under scenario 1).

*Scenario 3* is the same as scenario 1 with improved linkage to and retention in care: PWID are linked to care after an average duration of 0.5 years (vs. 2.6 years in scenario 1) and the loss to follow-up rate is 5%/year (vs. 13.8%/year in scenario 1).

*Scenario 4* is a combination of scenarios 2 and 3 (improved testing, linkage to and retention in care).

*Scenario 5* is scenario 1 with improved adherence to treatment. We considered that a better adherence could lead to a SVR rate similar to that of clinical trials, i.e.; $r=1$ (vs. 0.903 in scenario 1).

*Scenario 6* is scenario 1 with changes in treatment initiation criteria. All PWID who are tested, linked to care and who have no complications of cirrhosis are treated, regardless of fibrosis stage.

*Scenario 7* is the combination of scenarios 4, 5 and 6: improved testing, linkage to and retention in care; earlier treatment initiation criteria; and better adherence to treatment.

**Implementation of the model and outcomes**

For each scenario or sensitivity analysis, we performed 1,000 simulations. To ensure that results are comparable and to avoid the influence of the randomness in the network or population structure, we matched the simulations: each scenario was simulated on the same 1,000 simulated networks and PWID populations.

The impact of the scenarios on the prevalence and the incidence at 10 years, and on the difference in the number of new complications of cirrhosis in the population after 10 and 40 years was calculated and compared to the reference scenario. This time horizon for the number of cirrhosis complications was chosen because of the long delay before the occurrence of the complications in HCV infections.

**Sensitivity analysis**

We first performed a global deterministic sensitivity analysis to determine the parameters that have the most important impact under the reference scenario on the outcomes. We varied the values of the following parameters over the range of their uncertainty intervals, using data from other high income countries, or using estimates from expert opinion: the infection rates $\beta'$, the relative risk of infection during the first year of the injecting career, the testing rate $\delta$, and the linkage to care rate and lost to follow-up rate $\phi_{Link}$ and $\phi_{Lost}$, the transitions rates of the natural history model, and the initial distribution in the natural history model (Supplementary Information S5, Table S3).

We also performed specific sensitivity analysis about the parameters where uncertainty was important. First, when the parameter's estimate was uncertain because of the data source: the average number of injecting partners, which was fixed at 6 by hypothesis according to limited information from other countries (see Supplementary Information S1, Table S1). We also varied the values of the parameter for which the situation could change – or had already changed – compared to our estimates: the infection rate $\beta'$, as some evidence from different French sociological surveys show an increase in at-risk practices among PWID in France in recent years; and the risk of re-infection rate after a SVR, which can be higher than the primary infection rate (27, 28). Thus, we estimated the impact of a drop to an average of 3 partners or an increase to an average of 15 partners per PWID; and the impact of increasing the reference infection rate 5 ($\beta' = 0.05$) and 10 times ($\beta' = 0.1$) on all the simulated scenarios.

# RESULTS

**Impact of different interventions on the HCV-prevalence, incidence andrelated complications**

Figure 3 illustrates the impact of different interventions on the HCV infection prevalence and incidence at 10 years, and Figure 4 the impact on the number of complicationsof cirrhosis over 10 and 40 years. Tables S3 and S4in Supplementary Information present more details.

The HCV viral prevalence was set at 42.8% at the beginning of the simulation. With scenario 1 reflecting the current situation of HCV treatment, the mean prevalence decreased to 24.9% [95% confidence interval: 24.8%–24.9%] at 10 years. An improved testing performance (scenario 2) and/or linkage to care (scenario 3 and 4)or an improved adherence to treatment had a small impact on the results. Treating HCV-infected patients at the F0/F1 stage (scenario 6) decreased the prevalence at 10 years to 11.6% [11.6%–11.7%]. Finally, when combiningimproved testing, linkage to care, and adherence to treatment (scenario 7), we obtained a prevalence of 7% [7%–7.1%].Impact on the HCV incidence at 10 years followed comparable trends (see Figure 3).

Compared with the reference scenario (scenario 1),scenario 2, which includesimproved testing, had a small impact on the number of cirrhosis complications avoided after 10 and 40 years. Scenario 3, with an improved linkage to care, led to a decrease of9.8% [8.5%–11.1%] and 11.5% [10.0%–3.1%] over10 and 40 years respectively. In scenario 5, the improved adherence to treatment led toa decrease of3.2% [1.8%–4.6%] and 11.7% [10.0%–13.3%]. Treating infected PWID from F0/F1stage in scenario 6 did not have any impact on the number of cirrhosis complications. In scenario 7,with improved testing, linkage to care, adherence to treatment and early treatment the number of cirrhosis complications decrease was at 15.3% [14%–16.6%] and 29%; [27.9%;30.1%] over10 and 40 years respectively.

**Impact of different strategies on the cumulative number of treatments initiated**

The average cumulative number of treatment courses initiated during the first 10 years (Figure 5) is higher in scenarios 6 and 7, when HCV treatment is initiated early, with 3,978 and 4,066 treatments initiated, respectively,vs. scenarios 1 to 5, when treatment is initiated at liver fibrosis stages $\geq$F2, with2,349 and 2,404treatments initiated, respectively. In addition, the distribution of the number of treatments initiated overtime is different under different scenarios. Inscenarios 3, 4, 6, and 7, more treatments were used at the beginning of the simulation period compared to scenarios 1,2, and 5.

**Sensitivity analysis**

The detailed results of the deterministic sensitivity analysis of the model under the reference scenario are presented in Supplementary Information S5, Figure S2. The infection rate per partner, the transition rate between F0/F1 and F2/F3, the linkage to care/lost to follow-up rates are the parameters with the most important impact on HCV prevalence at 10 years, with variations of -3.1% to +4.7%, -2.2% to +2.6%, and -1.5% to +2.7%,respectively. The fourth parameter in order of importance is the average time to diagnosis, with variations of -0.2% to +2.5%. For the incidence at 10 years, the infection rate per partner, the average time to cessation and the relative risk of reinfection after a SVR are the most sensitive parameters with variations of respectively -0.6/100PY to +1.2/100PY, -0.2/100PY to +0.1/100PY and 0/100PY to 0.3/100PY. The transition rate between F2/F3 and F4, the initial fibrosis distribution and the linkage to care/loss to follow-up rates are the most sensitive parameters regarding to the cirrhosis complications over10 years.

In the other sensitivity analyses,results were robust to variations in the average number of injecting partners(3 or 15 vs. 6, see Table S4). When we increased the infection rate (see Figure S3 and S4)to $\beta' \times 5$ to take into account a possible increase in the PWID risk-taking behaviors,prevalence at 10 years was stable in the reference scenario, varying only from 42.8% to 43.7% [43.6%–43.8%] despite DAA use. When we increased the infection rate to $\beta' \times 10$, the prevalence at 10 years increased to 60.5% [60.4%–60.6%]. In these two cases, results were similar than in the basecase analysis.

# DISCUSSION

The proposed model allows comparing different scenarios impacting every step of the cascade of care (testing, linkage to care, treatment) on both the transmission of HCV and on the HCV-related morbidity. The results of the simulations showed several important points.

First, improving linkage to care or increasing the SVR rate to match that achieved with new DAAs decreased the number of cirrhosis complications by 10% on average after 40 years compared to the current situation. The benefit is relatively fast when improving linkage to care (9.8% after 10 years), meanwhile increasing the SVR rate had a more long-term impact. However, changing these parameters only had a small benefit on transmission, and, therefore, on the reduction in HCV prevalence (less than a 2% decrease in prevalence at 10 years). Second, the impact of improved testing was low on both transmission and morbidity. This trend is observed because, in France, HCV testing rate is already high: France has one of the highest rate of infection awareness in Europe(29, 30). In the sensitivity analysis, using testing rates observed in the UK, where the time between chronic infection and diagnosis is estimated to be 7.8 years (vs. 1.25/1.45 years in our model), we obtained for example a 10.4% increase of cirrhosis complications after 40 years. Thus, improved HCV testing could lead to a larger benefit in other settings. Third, initiating HCV treatment earlier, at F0/F1 fibrosis stage, had an important impact on transmission. The impact on morbidity was relatively moderate. From an individual perspective, treating early is not associated with an important benefit because a large proportion of the infected population never develop a liver complication (31). Those who do progress to later HCV fibrosis stages are usually detected and treated before developing complications using HCV fibrosis monitoring tools. Although, in our analysis we considered that we had perfect tools, which may be an optimistic hypothesis(32). In contrast, from a population perspective, treating patients early does prevent HCV transmission. Finally, improvements in testing, linkage to care, and adherence to treatment in addition to early treatment initiation allowed for a substantial decrease in both HCV transmission and morbidity.

The current recommendations suggest treating patients who reached F2, because HCV treatment regimens are costly and a high proportion of chronic hepatitis does not lead to complications(21).

Themodel shows that if the objective of policy-makers is the eradication of the virus, treating PWID from F0, in settings where the incidence remains high, is necessary.However, even in this case, eradication will be difficult to achieve and only important improvements on the cascade of care allowed decreasing the prevalence at 10 years below 10%.

The availability of a highly effective and well-tolerated treatment could potentially cause an increase in risky practices (i.e. needle sharing or some other example), as for HIV, for which the belief about highly active antiretroviral therapy had an impact on sexual risk behaviors, even if sexual behaviors and injection practices are not easily comparable due to different social practices (33). However, in our sensitivity analysis, we showed that our conclusions remained valid with an incidence 5 to 10 times greater than the base case.Since treatment regimens are expensive, the question of reinfection is an important one. In our model,using a re-infection rate per partner 10 times higher than the primary infection rate only had a low impact on the prevalence at 10 years, with less than 2% increase. However, the lack of data on the PWID social network in France constrained us to use a simple model with basic properties. Particularly, it shows no community structures (i.e.strongly linked and relatively isolated groups).Further data are needed to build a more sophisticated modelandclarify this important point.

This study presents several limitations. First, we placed our model in an ideal setting where the treatment regimens are given according to the national recommendations: PWID cannot decide or be told to reject treatment for any reason. In real life,because of the assumedrisk of reinfection and poor compliance of PWID there isresistance to treatment initiation.However, with the arrival of well-tolerated, injection-free treatments, with shorter durations, one may suppose that the treatment compliance will increase. Moreover, we hypothesized that there was only one line of HCV treatment and there was no retreatment after first treatment failure. We did not account for possible behavioral changes in PWID upon learning their hepatitis status (34), although they could impact the results.In this analysis, we did not limit the number of available treatment slots. Treating patients early (scenarios 6 and 7) sharply increased the cumulative amount of treatment used during the first years, which quickly reach 4,000 treatment courses after a few years, but remained stable thereafter.Given the current high costs associated with new DAAs (between 41,000 and 48,000 euros for a 12-week treatment course in France), budgetary impact will be an important issue especially when we consider treating patients early. In addition to costs, early treatment may have an important logistical and organizational impact for the medical system.

In conclusion we built an individual-based model taking into account the HCV infection, social network, natural history and cascade of care of chronic hepatitis C. We showed that to make the eradication of HCV possible, highly effective treatments are not sufficient in PWID, but an unconditional treatment, ideally associated with an improvement of access to care, is required. Several research pathways are offered by this model. In this context of costly treatment, a cost-effectiveness analysis should be considered in the future. In particular, we should evaluate the cost-effectiveness of strategies targeting PWID in the population according to their social network, that were shown to be more optimal in term of number of treatments needed(35). In addition, harm-reduction strategies such as supervised consumption rooms, improvement of needles/syringes exchange programs or substitution therapies, represent another way to impact the HCV epidemic. HCV treatment and care is not the only efficient strategy to decrease HCV transmission and the impact of these harm-reduction strategies along with treatment should be considered in the future to eradicate HCV infection.

## ACKNOWLEDGMENT

We would like to thank the scientific advisory board of this study for their helpful advices: Elisabeth Avril, PatriziaCarrieri, Elisabeth Delarocque-Astagneau, VéroniqueDoré, Albert Herszkowicz, Christine Larsen, Gilles Pialoux, Philippe Sogni, ElisabetaVergu. We also thank Margaret E. Hellardfor her helpful advicesand Marion Robine for linguistic revisions of the text and her advices. Numerical results presented in this paper were carried out using the regional computational cluster supported by Université Lille 1, CPER Nord-Pas-de-Calais/FEDER, France Grille, CNRS. We would like to thank the technical staff of the CRI-Lille 1 center.

# REFERENCES

1. WHO. Hepatitis C fact sheet n°164. In; 2013.
2. Shepard CW, Finelli L, Alter MJ. Global epidemiology of hepatitis C virus infection. Lancet Infect Dis 2005;5:558-567.
3. Nelson PK, Mathers BM, Cowie B, Hagan H, Des Jarlais D, Horyniak D, Degenhardt L. Global epidemiology of hepatitis B and hepatitis C in people who inject drugs: results of systematic reviews. Lancet 2011;378:571-583.
4. European Association for the Study of the Liver. EASL Clinical Practice Guidelines: management of hepatitis C virus infection. J Hepatol 2011;55:245-264.
5. Trepo C, Pradat P. Hepatitis C virus infection in Western Europe. J Hepatol 1999;31 Suppl 1:80-83.
6. Lawitz E, Mangia A, Wyles D, Rodriguez-Torres M, Hassanein T, Gordon S, Schultz M, et al. Sofosbuvir for previously untreated chronic hepatitis C infection. N Engl J Med 2013;368:1878-1887.
7. Afdhal N, Zeuzem S, Kwo P, Chojkier M, Gitlin N, Puoti M, Romero-Gomez M, et al. Ledipasvir and sofosbuvir for untreated HCV genotype 1 infection. N Engl J Med 2014;370:1889-1898.
8. Kowdley KV, Gordon SC, Reddy KR, Rossaro L, Bernstein DE, Lawitz E, Shiffman ML, et al. Ledipasvir and sofosbuvir for 8 or 12 weeks for chronic HCV without cirrhosis. N Engl J Med 2014;370:1879-1888.
9. Sulkowski MS, Gardiner DF, Rodriguez-Torres M, Reddy KR, Hassanein T, Jacobson I, Lawitz E, et al. Daclatasvir plus sofosbuvir for previously treated or untreated chronic HCV infection. N Engl J Med 2014;370:211-221.
10. Poordad F, Hezode C, Trinh R, Kowdley KV, Zeuzem S, Agarwal K, Shiffman ML, et al. ABT-450/r-ombitasvir and dasabuvir with ribavirin for hepatitis C with cirrhosis. N Engl J Med 2014;370:1973-1982.
11. Zeuzem S, Jacobson IM, Baykal T, Marinho RT, Poordad F, Bourliere M, Sulkowski MS, et al. Retreatment of HCV with ABT-450/r-ombitasvir and dasabuvir with ribavirin. N Engl J Med 2014;370:1604-1614.
12. Cousien A, Tran VC, Deuffic-Burban S, Jauffret-Roustide M, Dhersin JS, Yazdanpanah Y. Dynamic modelling of HCV transmission among drug users: a methodological review. Journal of Viral Hepatitis 2015;22:213-229.
13. Rolls DA, Daraganova G, Sacks-Davis R, Hellard M, Jenkinson R, McBryde E, Pattison PE, et al. Modelling hepatitis C transmission over a social network of injecting drug users. J Theor Biol 2011;297C:73-87.
14. Vaissade L, Legleye S. Capture-recapture estimates of the local prevalence of problem drug use in six French cities. Eur J Public Health 2009;19:32-37.
15. Lopez D, Martineau H, Palle C. Mortalité liée aux drogues illicites: OFDT, 2004.
16. Mathei C, Shkedy Z, Denis B, Kabali C, Aerts M, Molenberghs G, Van Damme P, et al. Evidence for a substantial role of sharing of injecting paraphernalia other than syringes/needles to the spread of hepatitis C among injecting drug users. J Viral Hepat 2006;13:560-570.
17. Newman MEJ. The structure and function of complex networks. SIAM 2003.
18. Erdős P, Rényi A. On Random Graphs. I. Publicationes Mathematicae 1959;6:290–297.
19. Terrault NA, Dodge JL, Murphy EL, Tavis JE, Kiss A, Levin TR, Gish RG, et al. Sexual transmission of hepatitis C virus among monogamous heterosexual couples: the HCV partners study. Hepatology 2013;57:881-889.
20. Sutton AJ, Gay NJ, Edmunds WJ, Hope VD, Gill ON, Hickman M. Modelling the force of infection for hepatitis B and hepatitis C in injecting drug users in England and Wales. BMC Infect Dis 2006;6:93.
21. Ministère des affaires sociales, de la santé et des droits des femmes. Arrêté du 30 octobre 2014 modifiant la liste des spécialités pharmaceutiques agréées à l'usage des collectivités et divers services publics. Journal Officiel de la République Française, n°255, 4 novembre 2014, texte 32. In; 2014.
22. The French METAVIR Cooperative Study Group. Intraobserver and interobserver variations in liver biopsy interpretation in patients with chronic hepatitis C. Hepatology 1994;20:15-20.

23.	van der Meer AJ, Veldt BJ, Feld JJ, Wedemeyer H, Dufour JF, Lammert F, Duarte-Rojo A, et al. Association between sustained virological response and all-cause mortality among patients with chronic hepatitis C and advanced hepatic fibrosis. JAMA 2012;308:2584-2593.
24.	Jauffret-Roustide M, Le Strat Y, Couturier E, Thierry D, Rondy M, Quaglia M, Razafandratsima N, et al. A national cross-sectional study among drug-users in France: epidemiology of HCV and highlight on practical and statistical aspects of the design. BMC Infect Dis 2009;9:113.
25.	Marin J-M, Pudlo P, Robert CP, Ryder RJ. Approximate Bayesian computational methods. Statistics and Computing 2012;22:1167-1180.
26.	Melin P, Chousterman M, Fontanges T, Ouzan D, Rotily M, Lang JP, Marcellin P, et al. Effectiveness of chronic hepatitis C treatment in drug users in routine clinical practice: results of a prospective cohort study. Eur J Gastroenterol Hepatol 2010;22:1050-1057.
27.	Aitken CK, Lewis J, Tracy SL, Spelman T, Bowden DS, Bharadwaj M, Drummer H, et al. High incidence of hepatitis C virus reinfection in a cohort of injecting drug users. Hepatology 2008;48:1746-1752.
28.	Micallef JM, Macdonald V, Jauncey M, Amin J, Rawlinson W, van Beek I, Kaldor JM, et al. High incidence of hepatitis C virus reinfection within a cohort of injecting drug users. J Viral Hepat 2007;14:413-418.
29.	Hutchinson SJ. Epidemiology & importance of monitoring. In: EASL - The International Liver Congress 2014, 49th Annual Meeting of the European Association for the Study of the Liver 2014 April 9-13; London, United-Kingdom; 2014.
30.	Cornberg M, Razavi HA, Alberti A, Bernasconi E, Buti M, Cooper C, Dalgard O, et al. A systematic review of hepatitis C virus epidemiology in Europe, Canada and Israel. Liver Int 2011;31 Suppl 2:30-60.
31.	Poynard T, Bedossa P, Opolon P. Natural history of liver fibrosis progression in patients with chronic hepatitis C. The OBSVIRC, METAVIR, CLINIVIR, and DOSVIRC groups. Lancet 1997;349:825-832.
32.	Shaheen AA, Wan AF, Myers RP. FibroTest and FibroScan for the prediction of hepatitis C-related fibrosis: a systematic review of diagnostic test accuracy. Am J Gastroenterol 2007;102:2589-2600.
33.	Crepaz N, Hart TA, Marks G. Highly active antiretroviral therapy and sexual risk behavior: a meta-analytic review. JAMA 2004;292:224-236.
34.	Aspinall EJ, Weir A, Sacks-Davis R, Spelman T, Grebely J, Higgs P, Hutchinson SJ, et al. Does informing people who inject drugs of their hepatitis C status influence their injecting behaviour? Analysis of the Networks II study. Int J Drug Policy 2014;25:179-182.
35.	Rolls DA, Sacks-Davis R, Jenkinson R, McBryde E, Pattison P, Robins G, Hellard M. Hepatitis C transmission and treatment in contact network of people who inject drugs. PLoS One 2013;8.

**Table 1** Description of the scenarios

| Scenario | | Average time to diagnosis (current/former PWID) $1/\delta$ | Average time to LTC $1/\phi_{Link}$ | Loss to follow-up rate $\phi_{Lost}$ | Treatment initiation criteria | %SVR $e$ |
|---|---|---|---|---|---|---|
| 1 (reference) | Incoming DAAs regimens | 1.25 y / 1.45 y | 2.6 y | 14%/y | F2 →F4 | 81.30% |
| 2 | Incoming DAAs regimens + improved testing | 0.5 y | 2.6 y | 14%/y | F2 →F4 | 81.30% |
| 3 | Incoming DAAs regimens + improved LTC | 1.25 y / 1.45 y | 0.5 y | 5%/y | F2 →F4 | 81.30% |
| 4 | Incoming DAAs regimens + improvement of testing + improvement of LTC | 0.5 y | 0.5 y | 5%/y | F2 →F4 | 81.30% |
| 5 | Incoming DAAs regimens + improved adherence to treatment | 1.25 y / 1.45 y | 2.6 y | 14%/y | F2 →F4 | 90.00% |
| 6 | Incoming DAAs regimens + earlier treatment initiation | 1.25 y / 1.45 y | 2.6 y | 14%/y | F0 →F4 | 81.30% |
| 7 | Incoming DAAs regimens + Changes in recommendations to treat earlier + improvement of testing + improvement of LTC + improvement of adherence to treatment | 0.5 y | 0.5 y | 5%/y | F0 →F4 | 90.00% |

SVR: Sustained virological response; PWID: People who inject drugs; DAA: Direct-acting antiviral; LTC: Linkage to care

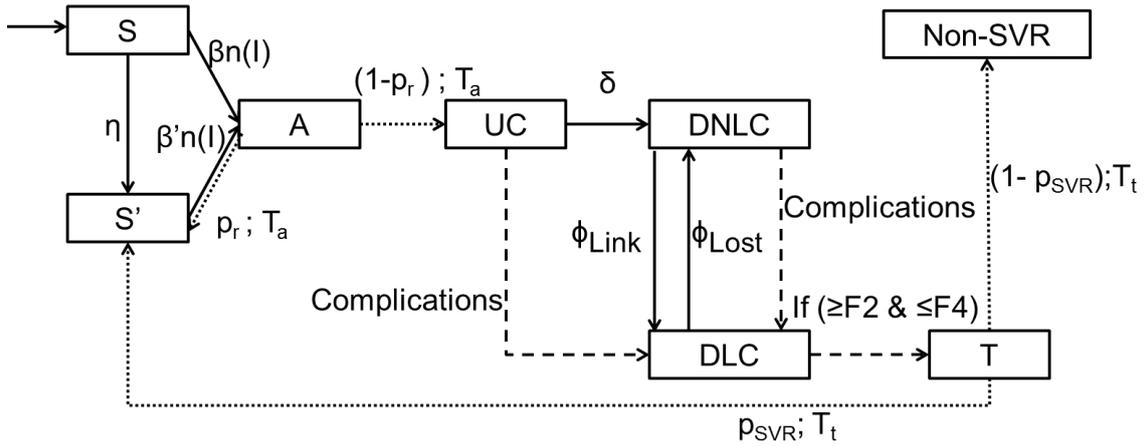

**Figure 1** Individual-based model for HCV infection and cascade of care. S: Susceptible people at high risk (inexperienced injector); S': Susceptible people at low risk (experienced injectors); A: Acute hepatitis C; UC: Undiagnosed chronic hepatitis C; DNLC: Diagnosed and non-linked to care with chronic hepatitis C; DLC: Diagnosed and linked to care with chronic hepatitis C; T: Treatment; Non-SVR: No sustained virological response. The Greek letters correspond to annual rates and the transitions occur according to exponential laws. n(I) is the number of current infected injecting partners of the PWID. The time spent with acute hepatitis C $T_a$ or on treatment $T_t$ is deterministic. The issue of these states, UC or S' in A, and Non-SVR or S' in T, is determined by a Bernoulli drawn set of parameter $p_r$ and $p_{SVR}$ respectively. In each state of the model, cessation of injections or death non-related to HCV occur according to exponential laws.

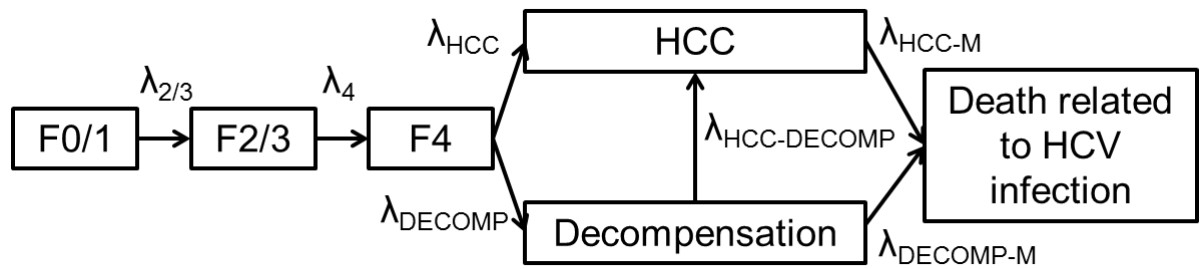

**Figure 2** Natural history model for chronic hepatitis C. We grouped Metavir score F0 and F1 (F0/1), and F2 and F3 (F2/3) for simplicity. The transitions times between the different states are drawn in exponential distributions.

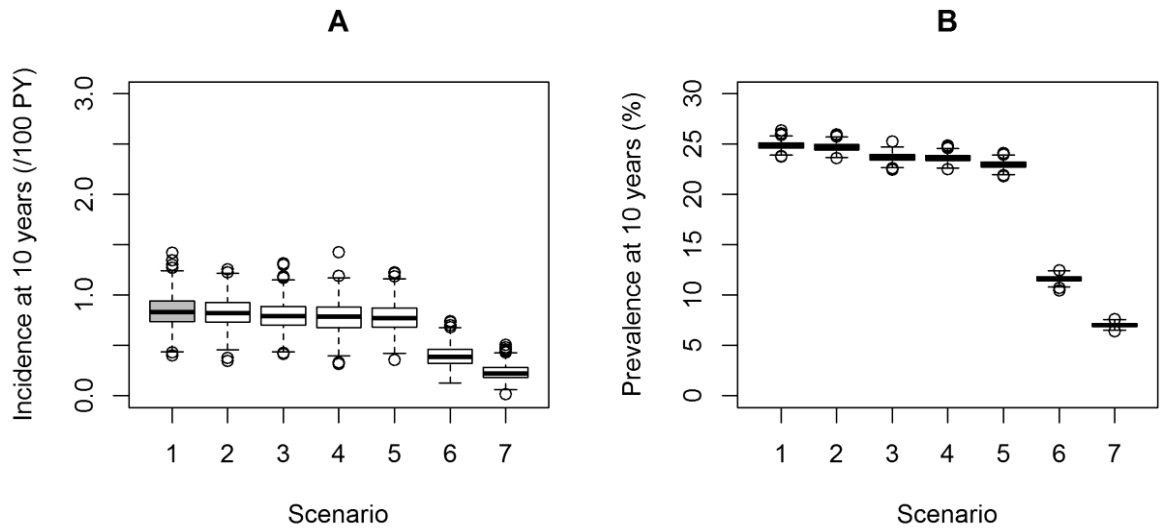

**Figure 3** Boxplots of the outcomes of the model according to the different scenarios. A. Prevalence at 10 years; B. Incidence at 10 years. The black line represents the median, the box represents the interquartile range. Whiskers maximum distance is 1.5 times the interquartile range.

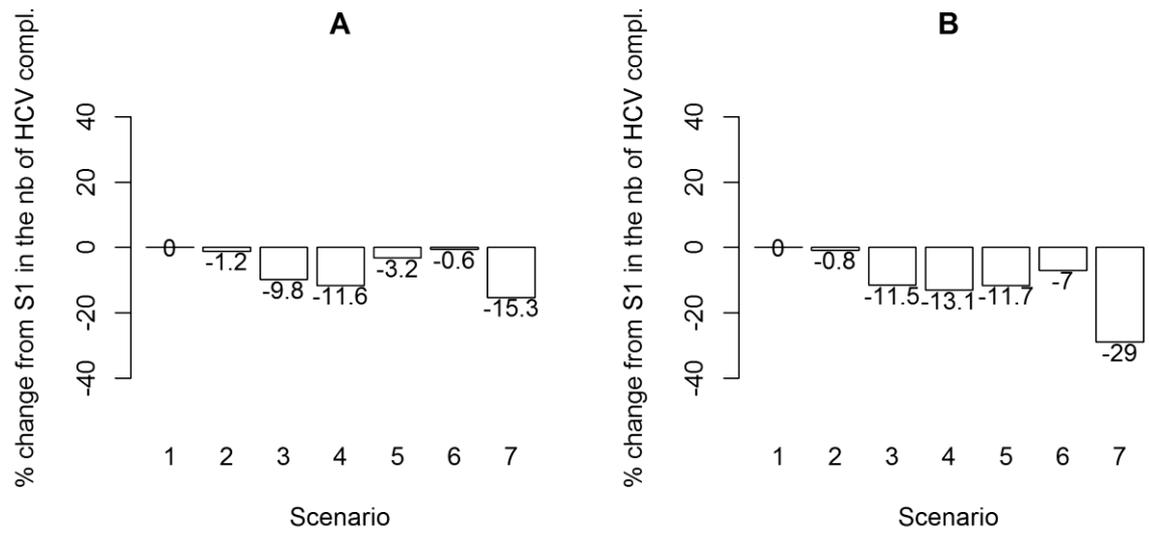

**Figure 4** Average percentage of cirrhosis complications avoided over A. 10 years and B. 40 years for each scenario, compared to scenario 1.

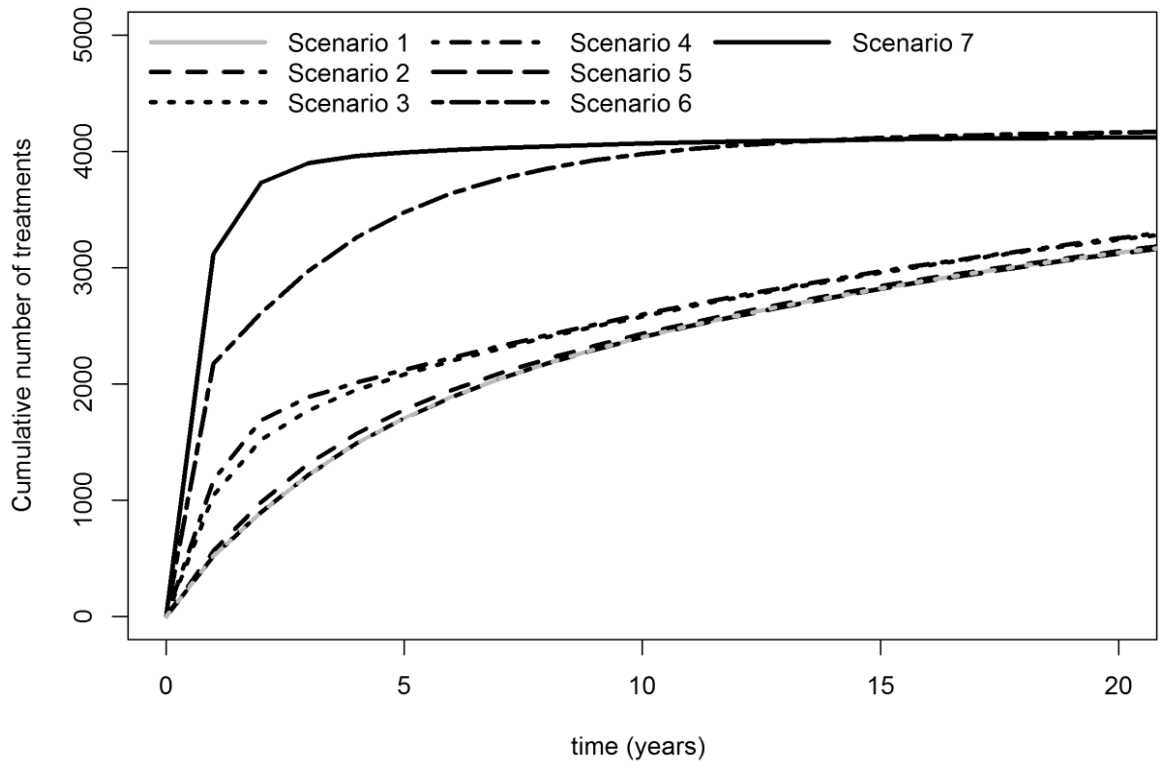

**Figure 5** Cumulative number of treatment courses initiated over the 10 years of simulations, for each scenario.